\input amstex
 \input epsf
\magnification=\magstep1 
\baselineskip=13pt
\documentstyle{amsppt}
\def \EE{\bold{E\thinspace}}
\def \per{\operatorname{per}}
\vsize=8.7truein \CenteredTagsOnSplits \NoRunningHeads

\topmatter
\title  A remark on approximating permanents of positive definite matrices \endtitle 
\author Alexander Barvinok  \endauthor
\address Department of Mathematics, University of Michigan, Ann Arbor,
MI 48109-1043, USA \endaddress
\email barvinok$\@$umich.edu  \endemail
\date May 12, 2020 \enddate
\thanks  This research was partially supported by NSF Grant DMS 1855428.
\endthanks 
\abstract Let $A$ be an $n \times n$ positive definite Hermitian matrix with all eigenvalues between 1 and 2. We represent the permanent of $A$ as the integral of some explicit log-concave function on ${\Bbb R}^{2n}$.
Consequently, there is a fully polynomial randomized approximation scheme (FPRAS) for $\per A$.
\endabstract
\endtopmatter

\document

\head 1. Introduction and main results \endhead

Let $A=\left(a_{ij}\right)$ be an $n \times n$ complex matrix. The {\it permanent} of $A$ is defined as 
$$\per A = \sum_{\sigma \in S_n} \prod_{k=1}^n a_{k \sigma(k)},$$
where $S_n$ is the symmetric group of all $n!$ permutations of the set $\{1, \ldots, n\}$. Recently, there was some interest in efficient computing (approximating) $\per A$, when $A$ is a positive definite Hermitian matrix (as is known, in that case $\per A$ is real and non-negative), see \cite{A+17} and reference therein. In particular, Anari et al. construct in \cite{A+17} a deterministic algorithm approximating the permanent of a positive semidefinite $n \times n$ Hermitian matrix $A$ within a multiplicative factor of $c^n$ for $c=e^{1+\gamma} \approx 4.84$, where $\gamma \approx 0.577$ is the Euler constant.

In this note, we show that that there is a fully polynomially randomized approximation scheme (FPRAS) for permanents of positive definite matrices with the eigenvalues between 1 and 2. Namely, we represent $\per A$ for such a matrix $A$ as an integral of an explicitly constructed log-concave function $f_A: {\Bbb R}^{2n} \longrightarrow {\Bbb R}$, so that a Markov Chain Monte Carlo algorithm can be applied to efficiently approximate 
$$\int_{{\Bbb R}^{2n}} f_A(x) \ dx =\per A,$$
see \cite{LV07}. 

We consider the space ${\Bbb C}^n$ with the standard norm
$$\|z\|^2=|z_1|^2 + \ldots + |z_n|^2, \quad \text{where} \quad z=\left(z_1, \ldots, z_n \right).$$
We identify ${\Bbb C}^n ={\Bbb R}^{2n}$ by identifying $z=x+iy$ with $(x, y)$. For a complex matrix $L=\left(l_{jk}\right)$, we denote by $L^{\ast}=\left(l^{\ast}_{jk}\right)$ its conjugate, so that 
$$l^{\ast}_{jk}=\overline{l_{kj}} \quad \text{for all} \quad j, k.$$

We prove the following main result.
\proclaim{(1.1) Theorem} Let $A$ be an $n \times n$ positive definite matrix with all eigenvalues between $1$ and $2$. Let us write $A=I +B$, where $I$ is the $n \times n$ identity matrix and $B$ is an $n \times n$ 
positive semidefinite Hermitian matrix with eigenvalues between $0$ and $1$. Further, we write $B=L L^{\ast}$, where $L=\left(l_{jk}\right)$ is an $n \times n$ complex matrix. We define linear functions 
$\ell_1, \ldots, \ell_n: {\Bbb C}^n \longrightarrow {\Bbb C}$ by 
$$\ell_j(z)=\sum_{k=1}^n l_{jk} z_k \quad \text{for} \quad z=\left(z_1, \ldots, z_n\right).$$
Let us define $f_A: {\Bbb C}^n \longrightarrow {\Bbb R}_+$ by 
$$ f_A(z)={1 \over \pi^n} e^{-\|z\|^2} \prod_{j=1}^n \left(1 + \left| \ell_j(z)\right|^2\right).$$
\roster
\item Identifying ${\Bbb C}^n={\Bbb R}^{2n}$, we have
$$\per A = \int_{{\Bbb R}^{2n}} f_A(x,y) \ dx dy.$$
\item The function $f_A: {\Bbb R}^{2n} \longrightarrow {\Bbb R}_+$ is log-concave, that is, 
$$\split &f_A\left(\alpha x_1 + (1-\alpha) x_2\right) \ \geq \ f_A^{\alpha}(x_1) f_A^{1-\alpha}(x_2) \\ & \text{for any} \quad x_1, x_2 \in {\Bbb R}^{2n} \quad \text{and any} \quad 0 \leq \alpha \leq 1.\endsplit $$
\endroster
\endproclaim

\head 2. Proofs \endhead

We start with a known integral representation of the permanent of a positive semidefinite matrix.
\subhead (2.1) The integral formula \endsubhead
Let $\mu$ be the Gaussian probability measure in ${\Bbb C}^n$ with density 
$${1 \over \pi^n} e^{-\|z\|^2} \quad \text{where} \quad \|z\|^2=|z_1|^2+ \ldots + |z_n|^2 \quad \text{for} \quad z=\left(z_1, \ldots, z_n \right).$$
Let $\ell_1, \ldots, \ell_n : {\Bbb C}^n \longrightarrow {\Bbb C}$ be linear functions and let $B=\left(b_{jk}\right)$ be the $n \times n$ matrix,
$$b_{jk}=\EE \ell_j \overline{\ell_k} = \int_{{\Bbb C}^n} \ell_j(z) \overline{\ell_k(z)} \ d \mu(z) \quad \text{for} \quad j,k=1, \ldots, n.$$
Hence $B$ is a Hermitian positive semidefinite matrix and the Wick formula (see, for example, Section 3.1.4 of \cite{Ba16}) implies that 
$$\per B = \EE \left( |\ell_1|^2 \cdots |\ell_n|^2\right) = \int_{{\Bbb C}^n} |\ell_1(z)|^2  \cdots |\ell_n(z)|^2 \ d \mu(z). \tag2.1.1$$

Next, we need a simple lemma.
\proclaim{(2.2) Lemma} Let $q: {\Bbb R}^n \longrightarrow {\Bbb R}_+$ be a positive semidefinite quadratic form. Then the function 
$$h(x) =\ln \bigl(1+q(x)\bigr) -q(x)$$
is concave.
\endproclaim
\demo{Proof} It suffices to check that the restriction of $h$ onto any affine line $x(\tau)=\tau a +b$ with $a, b \in {\Bbb R}^n$ is concave. Thus we need to check that the univariate function 
$$G(\tau)=\ln \bigl( 1+ (\alpha \tau + \beta)^2 +\gamma^2\bigr) - (\alpha \tau + \beta)^2 -\gamma^2  \quad \text{for} \quad \tau \in {\Bbb R},$$
where $\alpha \ne 0$, is concave, for which it suffices to check that $G''(\tau) \leq 0$ for all $\tau$. Via the affine substitution $\tau:=(\tau-\beta)/\alpha$, it suffices to check that 
$g''(\tau) \leq 0$, where
$$g(\tau)=\ln\left(1 + \tau^2 + \gamma^2\right) - \left(\tau^2+ \gamma^2\right).$$
We have
$$g'(\tau)={2 \tau \over 1+ \tau^2 +\gamma^2} - 2\tau$$
and
$$\split g''(\tau)=&{2 (1+\tau^2+\gamma^2) - 4\tau^2 \over \left(1+\tau^2 +\gamma^2\right)^2} -2 \\= &{2 (1+\tau^2+\gamma^2) - 4\tau^2 -2\left(1+\tau^2 +\gamma^2\right)^2 \over \left(1+\tau^2 +\gamma^2\right)^2}\\
=&{2 +2 \tau^2 +2 \gamma^2 -4\tau^2 -2 -2 \tau^4 -2 \gamma^4 -4 \tau^2 -4 \gamma^2 -4 \tau^2 \gamma^2 \over \left(1+\tau^2 +\gamma^2\right)^2} \\
=&-{6 \tau^2 +2 \gamma^2  +2 \tau^4 +2 \gamma^4  +4 \tau^2 \gamma^2 \over \left(1+\tau^2 +\gamma^2\right)^2} \leq 0
\endsplit$$
and the proof follows.
{\hfill \hfill \hfill} \qed
\enddemo

\subhead (2.3) Proof of Theorem 1.1 \endsubhead We have 
$$\per A=\per(I+B)=\sum_{J \subset \{1, \ldots, n\}} \per B_J,$$
where $B_J$ is the principal $|J| \times |J|$ submatrix of $B$ with row and column indices in $J$ and where we agree that $\per B _{\emptyset}=1$. Let us consider the Gaussian probability measure in ${\Bbb C}^n$ with density $\pi^{-n} e^{-\|z\|^2}$. By (2.1.1), we have 
$$\per B_J = \EE \prod_{j \in J} |\ell_j(z)|^2$$
and hence 
$$\per A = \EE \prod_{j=1}^n \left(1 + |\ell_j(z)|^2\right) =\int_{{\Bbb R}^{2n}} f_A(x, y) \ dx dy,$$
and the proof of Part (1) follows. 

We write
$$\split &e^{-\|z\|^2} \prod_{j=1}^n \left(1+|\ell_j(z)|^2 \right) = e^{-q(z)} \prod_{j=1}^n \left(1+ |\ell_j(z)|^2\right) e^{-|\ell_j(z)|^2}, \\ &\quad \text{where} \quad q(z)=\|z\|^2 - \sum_{j=1}^n |\ell_j(z)|^2.\endsplit$$
By Lemma 2.2 each function $(1+|\ell_j(z)|^2) e^{-|\ell_j(z)|^2}$ is log-concave on ${\Bbb R}^{2n} = {\Bbb C}^n$ and hence to complete the proof of Part (2) it suffices to show that $q$ is a positive semidefinite Hermitian form. To this end, we consider the Hermitian form
$$\split p(z)=&\sum_{j=1}^n |\ell_j(z)|^2=\sum_{j=1}^n \left| \sum_{k=1}^n l_{jk} z_k \right|^2 =\sum_{j=1}^n \sum_{1 \leq k_1, k_2 \leq n} l_{jk_1} \overline{l_{jk_2}} z_{k_1} \overline{z_{k_2}} \\
=&\sum_{1 \leq k_1, k_2 \leq n} c_{k_1 k_2} z_{k_1} \overline{z_{k_2}}, \endsplit$$
where 
$$c_{k_1 k_2} =\sum_{j=1}^n l_{j k_1} \overline{l_{j k_2}} \quad \text{for} \quad 1 \leq k_1, k_2 \leq n.$$
Hence for the matrix $C=\left(c_{k_1 k_2}\right)$ of $p$, we have $C=\overline{L^{\ast} L}$. We note that $B=L L^{\ast}$ and that the eigenvalues of $B$ lie between 0 and 1. Therefore, the eigenvalues of 
$L^{\ast} L$ lie between 0 and 1 (in the generic case, when $L$ is invertible, the matrices $L L^{\ast}$ and $L^{\ast} L$ are similar). Consequently, the eigenvalues of $C$ lie between $0$ and $1$ and hence the Hermitian form $q(z)$ with matrix $I- C$ is positive semidefinite, which completes the proof of Part (2).
{\hfill \hfill \hfill} \qed

\enddocument
\end